\documentclass[12pt]{article}
\usepackage{amsmath}
\usepackage{amsfonts}
\usepackage{amssymb}

\usepackage{graphicx}

\begin{document}

\title{Structure of trajectories of complex matrix eigenvalues 
in the Hermitian-nonHermitian transition}

\author{O. Bohigas$^{1}$, J. X. De Carvalho$^2$ and M. P. Pato$^{3}$\\
$^{1}$CNRS, Universit\'e Paris-Sud, UMR8626, \\
LPTMS, Orsay Cedex, F-91405, France \\
$^2$Universidade Federal do Rio de Janeiro, Xerem, 25250-000, \\
Duque de Caxias, R.J., Brazil\\
$^{3}$Inst\'{\i}tuto de F\'{\i}sica, Universidade de S\~{a}o Paulo\\
Caixa Postal 66318, 05314-970 S\~{a}o Paulo, S.P., Brazil}


\maketitle

\begin{abstract}
The statistical properties of trajectories of eigenvalues of Gaussian 
complex matrices whose  Hermitean condition is progressively broken 
are investigated. It is shown how the ordering on the real axis
of the real eigenvalues is reflected in the structure of the trajectories 
and also in the final distribution of the eigenvalues in the complex 
plane.   
\end{abstract}

\maketitle

\section{Introduction}

A few years after the introduction by E. Wigner\cite{Wigner} of the
Gaussian ensemble of Hermitean random matrices as a basis of a 
statistical theory of spectra of complex many-body systems,
Ginibre investigated properties of  the Gaussian matrices
when the Hermitian condition is removed\cite{Ginibre}. The Hermitian
ones have found applications\cite{Mitchell}, especially since the link 
with the characterization of the manifestation of chaos in quantum mechanics 
has been established\cite{Boh}. More recently, the nonHermitian 
class has attracted great interest\cite{Sommers}. 

The interpolation between universality classes of matrices is a 
standard subject of investigation in random matrix 
theories\cite{Hussein}. The Wigner-Ginibre transition, which is our 
present object of study, has already been matter of investigation. 
It has been established that
asymptotically, its density of eigenvalues, starting from Wigner's 
semi-circle law distribution
on the real axis, evolves as the transition proceeds through an 
elliptic shape, ending up in an uniform circular distribution at the 
Ginibre situation\cite{Gaudin}. In addition, an asymptotic weak 
non-Hermiticity regime\cite{Yan1} with important applications to quantum open 
systems\cite{Yan} has been found in the 
intermediate regime of transition.

From the physical point of
view, this transition can be seen as the passage of a gas from a 1D 
configuration to an isotropic 2D configuration. In fact, it is well
known that the positions of eigenvalues of Gaussian ensembles can be 
considered as those of charges of a Coulomb gas. It also has been 
recognized that the real eigenvalues of the Hermitian matrices have 
properties of a 1D gas of bosons, the so-called Girardeau 
gas\cite{Gira}. On the other hand, the joint distribution of the 
complex eigenvalues of the complete non-Hermitian matrices, that is, 
the Ginibre matrices have the same structure of the Laughlin 
wave function of a 2D gas\cite{Laughlin}.    

The behavior of individual eigenvalues of Hermitian random matrices
has been a matter of investigation since the pioneering works which 
described the distribution of largest eigenvalues, an achievement 
followed by several applications\cite{TW}. Recently, the order 
statistics problem of obtaining the distribution of all eigenvalues 
considered as an ordered sequence of random variables has been
addressed\cite{Pato}. Here we are interested in studying 
how that ordering on the real axis reflects on the eigenvalue 
trajectories in the complex plane as the  Hermitian condition is 
progressively removed and also in the final eigenvalue distribution. 
In other words, do the complex eigenvalues 
retain a memory of their initial positions on the real axis? To do this
investigation, we resort to a system of differential equations 
which describes the motion of the eigenvalues as a function of 
the parameter breaking the Hermitian condition.

\section{Transition equations and results}

The Ginibre ensemble consists of complex random matrices $S$  
whose joint distribution of elements is given by    
 
\begin{equation}
P (S )=\exp\left[- \mbox{tr}(S^{\dagger} S) \right] \label{1} 
\end{equation}
in which no Hermitian condition is imposed.
Specializing to the case we are interested, namely complex matrices
of size $N$, their eigenvalues, for large values of $N$, are uniformly 
distributed in a disk of radius $\sqrt{N}.$ Beyond this circle, at a 
radius $r$, the density decays as\cite{Mehta}   

\begin{equation}
\frac{\sqrt{\pi}\exp[-u^2]}{2u}
\end{equation}
where $u=r-\sqrt{N},$ such that eigenvalues can be found up to
$\sim \sqrt{N} + 2 .$ 

Taking a matrix out of this ensemble, we define a new matrix, $H(t),$ 
by the relation 

\begin{equation}
H(t) = \left(\frac{S+S^{\dagger}}{2}\right) +
t\left(\frac{S-S^{\dagger}}{2}\right) \label{2} .
\end{equation}
where the parameter $t$ varies from zero to one. It is relevant
to mention that others' parametrizations have been used to study this
transition\cite{Yan}; however, for our purpose of investigating 
trajectories, the parameter $t$ defined by Eq. (\ref{2}) is more 
convenient. Of course, our $t$ is related to those others parameters 
by a simple transformation. With $t>0,$ Eq. (\ref{2}) together 
with its adjoint, can be inverted to express $S$ in terms of $H$ 
and $H^{\dagger}$ as

\begin{equation}
S = \left(\frac{1+t}{2t}\right) H  
-\left(\frac{1-t}{2t}\right)H^{\dagger} . \label{3} 
\end{equation}  
Substituting  (\ref{3}) in (\ref{1}), we obtain
the density distribution of the $t-$dependent
matrix elements of $H$

\begin{equation}
P (H )=K_N (t) \exp\left(- \mbox{tr} \left[ \frac{1+t^2}{2t^2}
(H^{\dagger} H) - \frac{1-t^2}{4t^2}( HH
+H^{\dagger}H^{\dagger})\right]  
\right) .    \label{1b} 
\end{equation}
For $t=0$ the matrices $H(0)$ are  Hermitian such that, in the limit
$t\rightarrow 0,$  (\ref{1b})
becomes
    
\begin{equation}
P [H(0)]=\exp\left(- \mbox{tr}\left[H^{2}(0) \right] \right)  .
\end{equation}
Therefore $H(0)$ belongs to the Gaussian Unitary Ensemble whose
eigenvalues are distributed on the real axis according to the Wigner
semi-circle law\cite{Mehta}
 
\begin{equation}
\rho(x )= \frac{1}{\pi}\sqrt{2N-x^2} .
\end{equation} 
The joint distribution of the eigenvalues apart from a constant
is given by

\begin{equation}
P (z_1,z_2,...,z_N )= \exp\left[- \sum^{N}_{k=1}
\left( x_{k}^2 + \frac{y^2}{t^2}\right) \right] \prod_{j>i} 
\mid z_j -  z_i \mid ^2  =\exp\left[-W(z_1,...,z_N)\right] , 
\end{equation}
where

\begin{equation}
W(z_1,...,z_N)=  \sum^{N}_{k=1}
\left( x_{k}^2 + \frac{y^2}{t^2}\right)-2 \sum_{j>i}\log 
\mid z_j -  z_i \mid   .
\end{equation}
Therefore, the eigenvalues can be considered as positions of $N$ point
charges under the action of a confining potential harmonic in the
two Cartesian axis and a repulsive 1D Coulomb force.  

For intermediate values of t, that is $0< t <1,$ asymptotically,
that is when $N$ becomes large, the eigenvalues fill an 
ellipse\cite{Gaudin} whose axes for our parameterization are

\begin{equation}
a=\sqrt{\frac{2N}{1+t^2}}
\end{equation}
and

\begin{equation}
b=t^2\sqrt{\frac{2N}{1+t^2}} .
\end{equation}

As $S=H(1)$ 
belongs to the Ginibre ensemble of non-Hermitian matrices, 
$H(t)$ undergoes a transition from the Wigner to the 
Ginibre ensemble with eigenvalues moving along trajectories 
in the complex plane. Their motion is governed by a system of
differential equations which can be deduced in the following 
way\cite{Pato1}. The matrices $H(t)$ are  diagonalized by the 
similarity transformation 

\begin{equation}
D=Q^{-1}HQ  ,     \label{51a}
\end{equation}
where $D$ is a diagonal matrix whose diagonal contains the complex
eigenvalues while $Q$ is a matrix of size $N$ which contains the
eigenvectors. On the hand, the adjoint of Eq. (\ref{31})

\begin{equation}
D^{\dagger}=Q^{\dagger}H^{\dagger}\left(Q^{-1}\right)^{\dagger} \label{51b}
\end{equation}
shows that the inverse of the adjoint of $Q$ diagonalizes the Hermitian of $H.$ 
All these matrices are functions of the parameter $t.$
Taking the derivative (denoted by a dot) with respect to $t$ of 
(\ref{51a}) we obtain

\begin{equation}
\dot D =[D,U] + P, \label{31}
\end{equation}
where 

\begin{equation}
P=Q^{-1}\dot H Q
\end{equation}
and 

\begin{equation}
U=Q^{-1}\dot Q = - \dot Q^{-1} Q   \label{31f} 
\end{equation}
while for the derivative of $P$ we have

\begin{equation}
\dot P =[P,U] + Q^{-1}\ddot H Q .     \label{31d}   
\end{equation}
The diagonal part of (\ref{31}) gives

\begin{equation}
\dot D _{kk}=\dot z_{k}=P_{kk}, \label{31a}
\end{equation}
while the off-diagonal part 

\begin{equation}
\dot D_{kl}=0=D_{kk}U_{kl}-U_{kl}D_{ll}+P_{kl} \label{31b}
\end{equation}
yields

\begin{equation}
U_{kl}=-\frac{P_{kl}}{z_k-z_l} .       \label{31c}     
\end{equation}
Using the arbitrariness of the matrix $U$ we impose
the necessary condition that its diagonal elements vanish, 
that is $U_{kk}=0.$  Finally, from (\ref{31f}) we derive
the equations

\begin{equation}
\dot Q_{ij} =\sum_{l\ne j}\frac{Q_{kl}P_{lj}}{z_l - z_l }     
\end{equation}
and 

\begin{equation}
 \dot Q^{-1}_{ij} =\sum_{l\ne i}\frac{P_{il} Q^{-1}_{lj}}{z_i - z_l }        
\end{equation}
for the evolution of the eigenvectors.
The above equations together with the initial 
conditions provided by the Hermitian matrix $H(0)$ form a complete 
system of first order differential equations which numerically 
can be solved to obtain eigenvalues and eigenvectors along the transition. 

Insofar as the dependence of $H$ with the parameter $t$ has not been 
specified, this set of differential equations is general. 
Assuming this dependence to be given 
by (\ref{2}), the second derivative  $\ddot{H}$ 
vanishes due to the linear dependence with $t$ and, explicitly,
Eq. (\ref{31d}) gives

\begin{equation}
\dot P_{kk} =\sum_{m\ne k}\frac{2P_{km}P_{mk}}{z_k - z_m }      \label{33a}   
\end{equation}
for the diagonal elements and 

\begin{equation}
\dot P_{kl} =\sum_{m\ne k,m\ne l}P_{km}P_{mk}\left(\frac{1}{z_k - z_m}+
\frac{1}{z_l - z_m}\right)      \label{33b}   
\end{equation}
for the others elements. These two above equations together with
Eq. (\ref{31b}) form a set of $2N(N+1)$ coupled differential equations
to be integrated. Regarding the initial conditions, since $H(0)$ is 
Hermitian, its eigenvalues are real and its
eigenvector matrix is unitary, that is $Q^{-1}=Q^{\dagger}.$ Taking
this into account, we find that initially the matrix $P$ is given
by

\begin{equation}
P(0)=Q^{\dagger}(0)\frac{S-S^{\dagger}}{2}  Q(0) ,
\end{equation}
such that  

\begin{equation}
P(0)=-P^{\dagger}(0) ,
\end{equation}
whose diagonal part 

\begin{equation}
P_{kk}(0)=-P_{kk}^{*}(0) 
\end{equation}
shows that the diagonal elements of $P(0)$ are pure
imaginary. Substituting these initial `velocities' in (\ref{31a}),
we find that the eigenvalues leave the real axis  at $t=0$ 
perpendicularly. 

We observe that integrating these equations 
between the initial value $t=0$ and some final value $t=t_f$ is
equivalent to diagonalize $H(t)$ at each intermediate value of $t.$
This is illustrated in Fig. 1, in which $N=20$ eigenvalue trajectories 
obtained performing the integration from $t=0$ to $t=1$ the equations
of motion are compared with the result of diagonalizing the matrices 
at several intermediate values of $t.$ The agreement is
perfect showing that the above system is reliable and amenable to 
numerical integration.

Fig. 1 also shows that eigenvalues may have complicated trajectories in
the complex plane. Therefore, in order to reveal structures the trajectories
may have, it is necessary to accumulate results of many
simulations, that is to construct an ensemble. This is done in
Fig. 2 in which trajectories were obtained by evolving eigenvalues of
60 different initial Hermitian matrices, clearly exhibiting a structure.   
Indeed, trajectories starting on the real axis at the edge, cover regions of a
meniscus shape whose curvature decreases as eigenvalues more to the center are
considered. Trajectories starting at the central region of the
initial spectrum move inside strips. The presence of these strips is better
seen in Fig. 3 where points correspond to eigenvalues at the end of the
transition, that is for $t=1.$ The structure of the distribution of
points show that, statistically, eigenvalues preserve to some extent
the relative positions they had on the real axis.   

\section{Concluding remarks}

In conclusion, evolving under the action of the external confining harmonic potential
and the repulsion force among them, eigenvalues present structures in their 
trajectories and in their final distribution in the
complex plane which reflect the ordering they have on the real axis.
We remark that the present study can be considered as an instance of the so-called
parametric statistics used to characterize the evolution of individual 
eigenvalues as a function of an external parameter\cite{Simons}. 
In the case of complex eigenvalues, parametric evolution has been 
experimentally investigated by considering resonances trapped 
in an open microwave cavity in which the slit width can be 
varied\cite{Persson}.
  

This work is supported by a CAPES/COFECUB project and MPP is supported
by the Brazilian agencies CNPq and FAPESP.


{\bf Figure Captions}

Fig. 1 The lines are $N=20$ eigenvalue trajectories generated
integrating the equations of motion, and the dots over
the lines are eigenvalues obtained by diagonalizing the matrix.

Fig. 2 Ensemble of trajectories of 60 matrices of size $N=20$
generated evolving eigenvalues initially at the edges, the middle
parts, and the center of the spectra, together with the density
circle of radius $\sqrt{N}.$ 

Fig. 3 Ensemble of eigenvalues at the Ginibre regime evolved from
eigenvalues at the edges, the middle parts, and the center of the 
spectra, together with the density circle of radius $\sqrt{N}.$ 

\begin{figure}[p]
\includegraphics{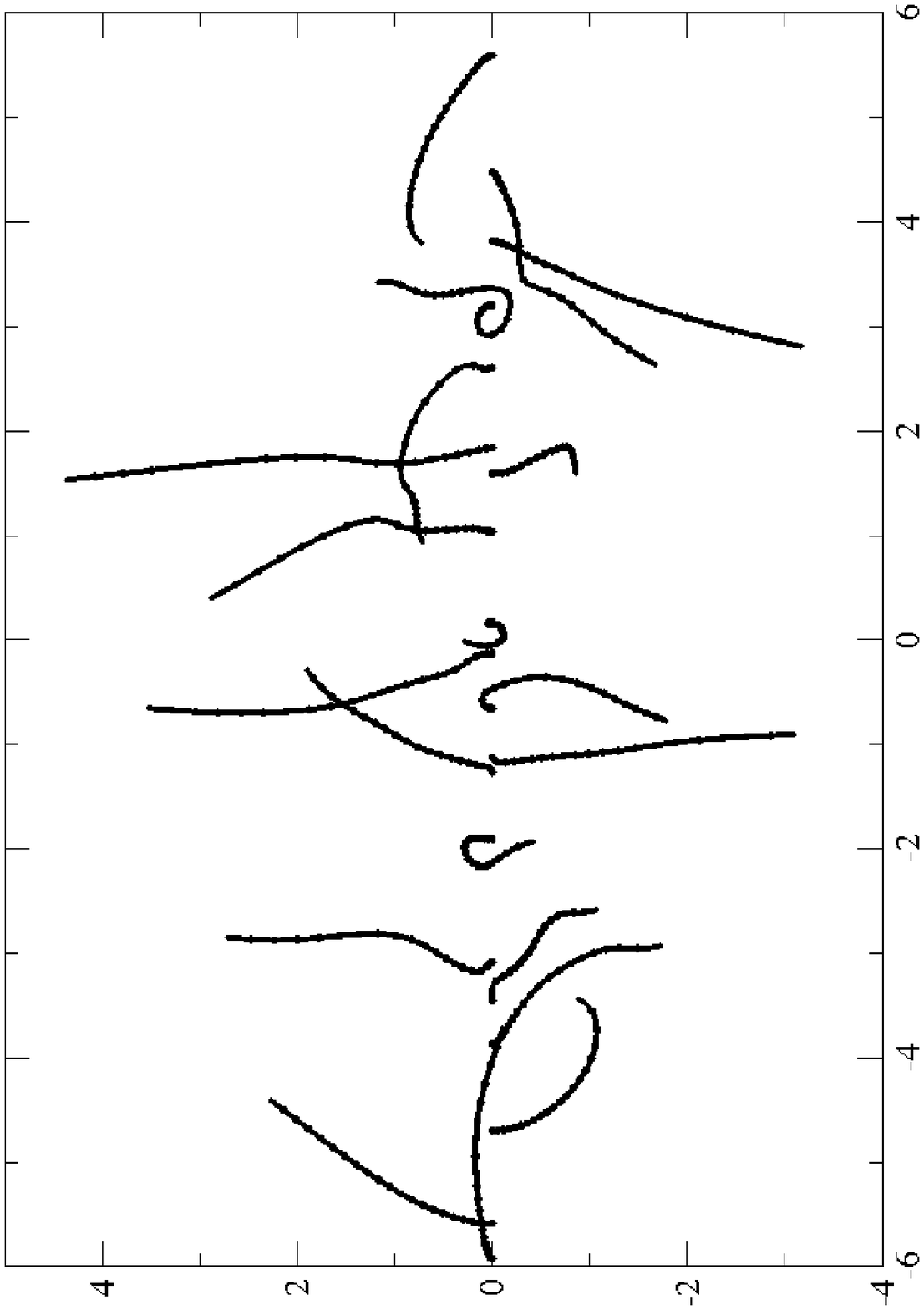}
\end{figure}

\begin{figure}[p]
\includegraphics{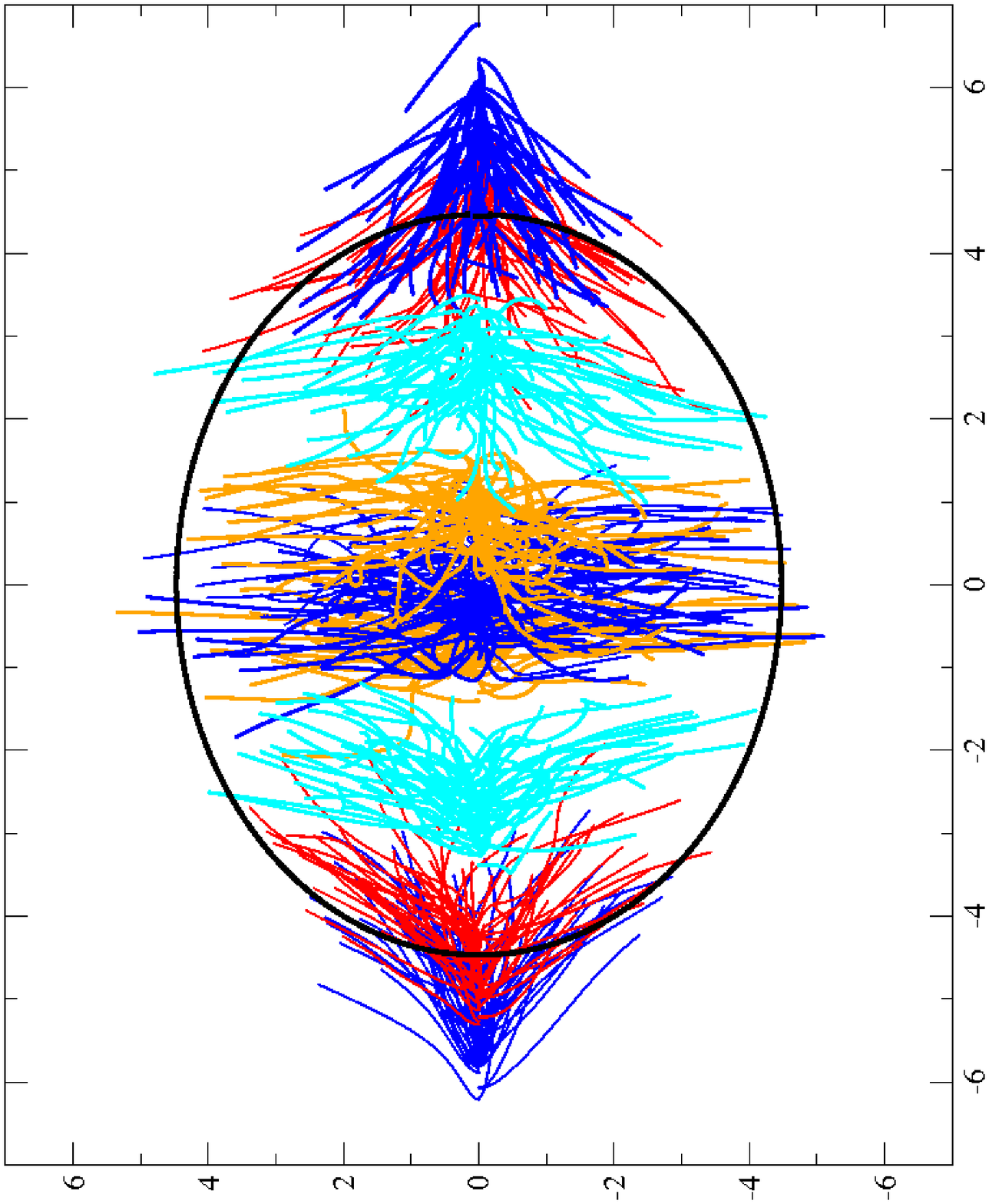}
\end{figure}

\begin{figure}[p]
\includegraphics{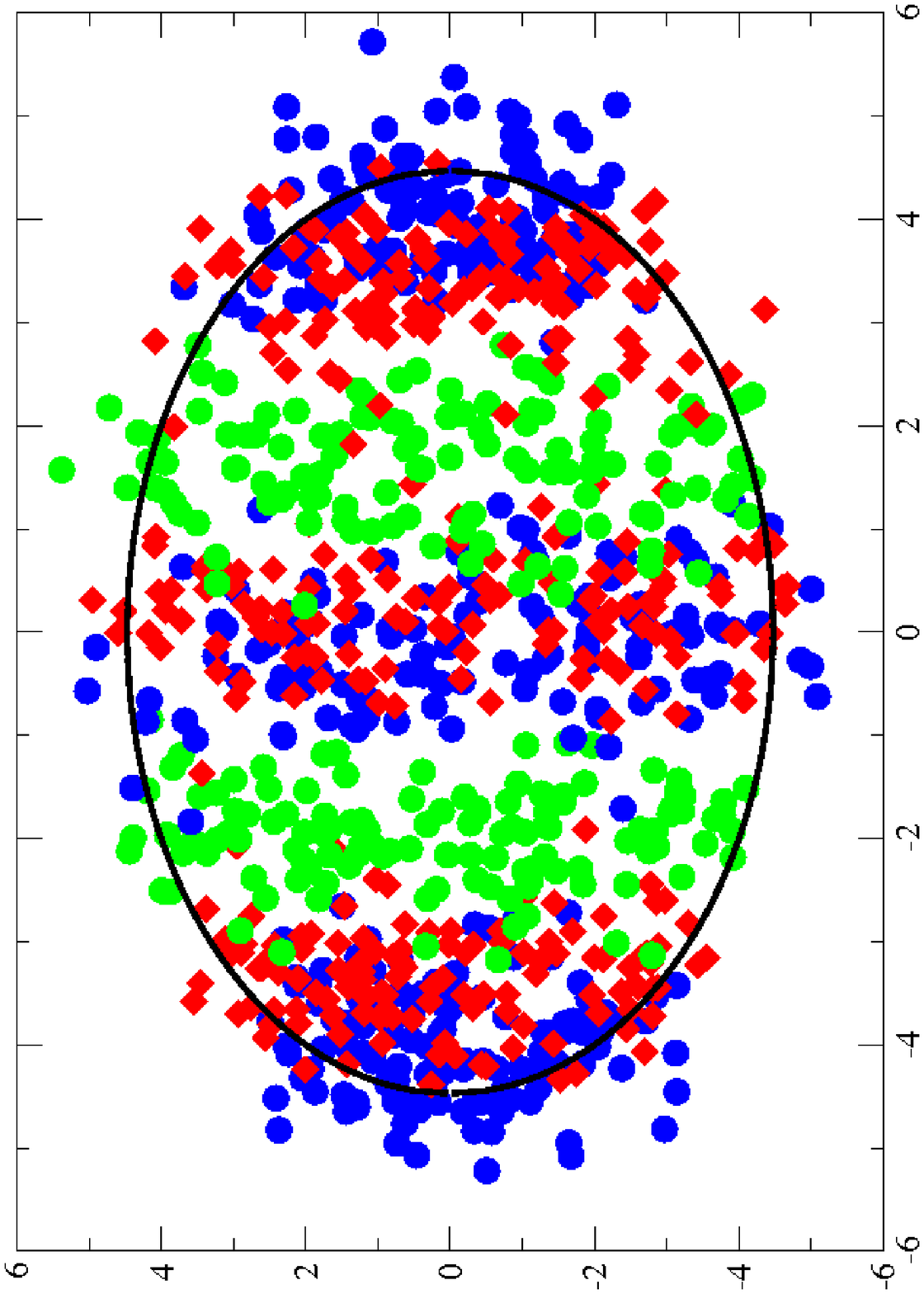}
\end{figure}

\end{document}